\documentclass[%
preprint,
 amsmath,amssymb,
 aps,
]{revtex4-2}

\usepackage{graphicx}
\usepackage{dcolumn}
\usepackage{bm}
\usepackage[utf8]{inputenc}
\usepackage{xcolor}
\usepackage{hyperref}

\begin{document}

\preprint{APS/123-QED}

\title{Stokes--anti-Stokes correlations of light propagating through weakly guiding optical fiber}
\author{Ivan V. Panyukov}
 \affiliation{Dukhov Research Institute of Automatics (VNIIA), 22 Sushchevskaya, Moscow 127055, Russia;}
 \affiliation{Moscow Institute of Physics and Technology, 9 Institutskiy pereulok, Dolgoprudny 141700, Moscow region, Russia;}
\author{Evgeny S. Andrianov}
 \affiliation{Dukhov Research Institute of Automatics (VNIIA), 22 Sushchevskaya, Moscow 127055, Russia;}
 \affiliation{Moscow Institute of Physics and Technology, 9 Institutskiy pereulok, Dolgoprudny 141700, Moscow region, Russia;}

\date{\today}

\begin{abstract}
Statistical properties of light produced in spontaneous Raman scattering on an ensemble of molecules indicate the quantum nature of this phenomenon.
The scattered light is non-classical and has high non-classical intensity correlations between Stokes and anti-Stokes components.
The temporal coherence of this light is well investigated, while many questions related to spatial coherence remain open.
Recent experiments reveal two peculiar features of the spatial coherence of the Stokes and anti-Stokes light.
First, the intensity correlations between Stokes and anti-Stokes light remain non-classical even for macroscopic samples containing many molecules.
Second, these correlations decrease when signal propagates through a multi-mode optical fiber: the more propagating fiber modes at Stokes and anti-Stokes frequencies the less the correlations.
Moreover, the second-order autocorrelation function of Stokes and anti-Stokes light also decreases with the number of propagating modes in multi-mode fiber.
In this paper, we build a model of spontaneous Raman scattering correlations of light produced by an ensemble of molecules and propagating through weakly guiding optical fiber that quantitatively explains all these observations.
We show that spacial orthogonality of the fiber modes makes the light propagating through these modes uncorrelated in the standard detection scheme.
This leads to suppression of non-classical intensity correlations of the total field in the multi-mode fiber.
We find the degree of non-classical correlations on fiber parameters.
The obtained results pave the way for engineering of non-classical Stokes -- anti-Stokes correlations.
\end{abstract}

\maketitle

\section{Introduction}

Stokes and anti-Stokes components of Raman light scattered on an ensemble of molecules are non-classically correlated~\cite{kasperczyk2016temporal, vento2023measurement, saraiva2017photonic, bustard2015nonclassical}.
The high frequencies of the molecular vibrations ($\sim 10$~THz) allow the observation of these non-classical correlations at room temperatures.
Before the Stokes scattering event, the molecular vibrations are almost in the ground state, suppressing the probability of anti-Stokes scattering.
Stokes scattering excites the molecular vibrations, which increases the probability of anti-Stokes scattering and causes the formation of non-classical correlations between Stokes and anti-Stokes light.
The list of the systems capable of producing light with non-classical correlations between the Stokes and the anti-Stokes components goes beyond ensembles of molecules and includes diamonds~\cite{kasperczyk2015stokes, velez2019preparation}, micromechanical systems~\cite{riedinger2016non, hong2017hanbury, galland2014heralded, anderson2018two, marinkovic2018optomechanical}, and atomic ensembles~\cite{duan2001long, chou2004single, farrera2016generation, park2018experimental, zhang2012coherent, corzo2019waveguide}.
Non-classical correlation in Raman light can be useful for quantum information processing~\cite{lee2012macroscopic, england2013photons, england2015storage, hou2016quantum, fisher2016frequency, fisher2017storage} and generation of non-classical states of light~\cite{tarrago2020bell}.

\begin{figure} 
\includegraphics[width=1\linewidth]{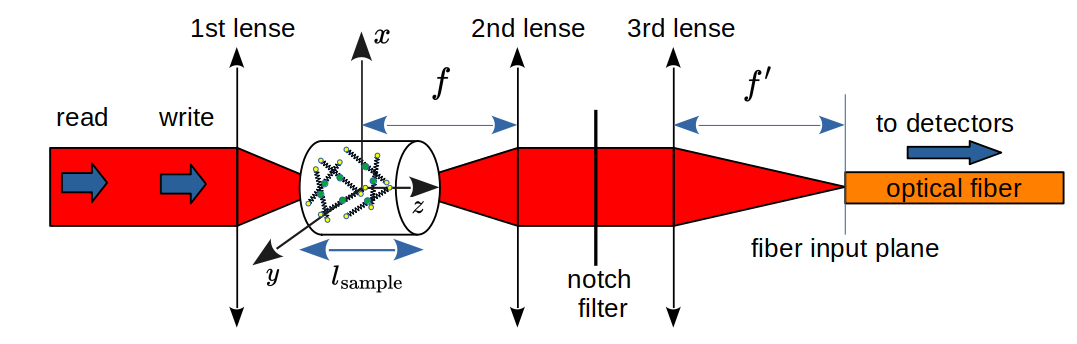}
\caption{
Optical setup for the measurement of Stokes--anti-Stokes correlations. 
The sample is in the focus of the first lens.
Light arising from the spontaneous Raman scattering on the molecules of the sample propagates through the optical system and the fiber to the detectors. 
The notch filter removes the write and read pulses. 
} \label{setup}
\end{figure}

In the experiments on the non-classical correlations in Raman scattering~\cite{anderson2018two, vento2023measurement} two spectrally separated pulses, {\it write} pulse and {\it read} pulse, induce the scattered light.
Then, the frequency filters select particular spectral components for the detection: the Stokes spectral component for the write pulse and the anti-Stokes spectral component for the read pulse.
The spectral separation of the read and write pulses simplifies the measurement of Stokes and anti-Stokes light.
Regarding the standard experimental setup, the molecules are in the focus of a lens that concentrates the intensity of the read and write pulses.
Next, two lenses focus the scattered light on the input of an optical fiber ({Fig.~\ref{setup}}).
The scattered light propagates through the optical fiber to the detectors that measure the second-order cross-correlation function of the Stokes and anti-Stokes light.

Non-classically correlated Stokes and anti-Stokes light have particular properties of polarization, temporal, and spatial coherence.
Both Stokes and anti-Stokes light inherits the polarization of the excitation laser~\cite{junior2020lifetime}.
The lifetime of the vibrational states~\cite{velez2019preparation} bounds the temporal coherence of the correlations.
The macroscopic number of molecules in the eliminated area can share the Stokes--anti-Stokes correlations, which may cause the oscillations of the second-order cross-correlation function~\cite{vento2023measurement}.
The measuring setup influences the Stokes and anti-Stokes light correlations~\cite{vento2023measurement}.
In particular, the optical fiber through which the signals propagate to the detectors affects the measured second-order cross-correlation and autocorrelation functions.
Increasing the number of optical fiber modes decreases the measured correlations~\cite{vento2023measurement}.
The many-body nature of the problem is the main obstacle to quantitatively describing this phenomenon.
This many-body problem includes the macroscopic number of molecules and several propagating modes in the waveguide. 
Solving this problem would enhance our understanding of the propagation of quantum correlations carried by light.

In this paper, we study non-classical Stokes--anti-Stokes correlations of the light scattered on an ensemble of molecules and propagating through a weakly guiding optical fiber.
We use the Heisenberg--Langevin equations~\cite{gardiner2004quantum} and obtain the electric field operators of Stokes and anti-Stokes signals.
We calculate the second-order cross-correlation function and second-order autocorrelation functions in the limit of a weakly guiding fiber and give a quantitative description of the decreasing of the correlations with the number of propagating modes in a waveguide, observed in~\cite{vento2023measurement}.
We also reproduce the quantum beats in the second-order cross-correlation function, observed in~\cite{vento2023measurement}, and theoretically confirm that frequencies of the beats are defined by the differences between frequencies of the molecular vibrations hosted by different molecules.
We show that quantum amplitudes corresponding to excitation of different fiber modes do not correlate due to their spacial orthogonality. 
It leads to suppression of the quantum intensity correlations of the light propagating through the fiber.
We explore how this suppression depends on the fiber parameters.

\section{Spontaneous Raman scattering}
We consider an ensemble of $N^{\rm (mol)}$ non-interacting molecules illuminated with an external light. 
These molecules are not necessarily the same.
If the sample contains several different types of molecules, we assume that the number of molecules of each type is macroscopic.
Each molecule hosts electronic excitation, which we model as a two-level system (TLS), and vibrational states, which we model as harmonic oscillators.
The electronic states have a transition dipole moment, allowing the direct coupling with the external light source.
The frequency discrepancy between the vibrational states and the external light source suppresses their direct interaction.
Thus, we omit the interaction between the external source and molecular vibrations.
Nevertheless, the coupling between the electronic and vibrational states of the molecules strongly affects the scattered light due to the non-linear nature of this interaction. 
Hamiltonian of the molecules interacting with the external field is
\begin{equation}
\hat H = \sum_{m=1}^{N^{(\rm mol)}} \hat H^{(\rm mol)}_m + \hat H^{\rm (drive)}(t),
\end{equation}
where
\begin{equation} \label{Hamiltonian_molecule}
\hat H^{(\rm mol)}_m = 
\hbar
\bigg[
\omega^{(\rm el)}_m 
+
\sum_{j=1}^{N^{\rm (vib)}_{m}} \Lambda_{mj}\omega^{(\rm vib)}_{mj} (\hat b^{\dag}_{mj} + \hat b_{mj})
\bigg]
\hat \sigma^{\dag}_m \hat\sigma_m +
\sum_{j=1}^{N^{\rm (vib)}_{m}} \hbar \omega^{(\rm vib)}_{mj} \hat b^{\dag}_{mj} \hat b_{mj},
\end{equation}
\begin{multline}
\hat H^{\rm (drive)}(t) =
\sum_{m=1}^{N^{\rm (mol)}}
\hbar 
\bigg[
\Omega_{m}^{\rm (write)}(t) \hat \sigma^{\dag}_m + \Omega^{{\rm (write)}*}_{m}(t) \hat\sigma_m
\bigg]
+ \\
\sum_{m=1}^{N^{\rm (mol)}}
\hbar
\bigg[
\Omega_{m}^{\rm (read)}(t-\tau) \hat \sigma^{\dag}_m + \Omega^{{\rm (read)}*}_{m}(t-\tau) \hat\sigma_m
\bigg]
,
\end{multline}
where $\omega^{(\rm el)}_m$ is the transition frequency, $\hat{\sigma}_{m}$ is the lowering operator of TLS.
Naturally, we assume that $\omega^{(\rm el)}_m$ are in the visible or ultraviolet range.
We represent the vibrational subsystem of $m$th molecule as a set of $N^{\rm (vib)}_m$ oscillators with the annihilation operator of $j$th modes $\hat{b}_{mj}$ and the corresponding frequency $\omega^{(\rm vib)}_{mj}$.
Thus, we consider the case of, in general, many vibrational modes in each molecule.
We assume that $\omega^{(\rm vib)}_{mj}/2\pi\sim 10$ THz, thus $\omega^{(\rm vib)}_{mj} \ll \omega^{(\rm el)}_m$.
Huang--Rhys factor, $\Lambda_{mj}^2$, characterizes the interaction between the electronic states of $m$th molecule with $j$th vibrational mode of this molecule.
The constants ${\Omega}_{m}^{\rm (write)}(t)=-{\bf d}_m {{\bf E}^{(\rm write)}}({\bf R}_m,t)/\hbar$ and ${\Omega}_{m}^{\rm (read)}(t)=-{\bf d}_m {{\bf E}^{(\rm read)}}({\bf R}_m,t)/\hbar$ characterize the interaction between the electronic states of $m$th molecule and read or write pulses having electric fields ${{\bf E}^{(\rm write)}}({\bf R}_m,t)$ and ${{\bf E}^{(\rm read)}}({\bf R}_m,t)$ respectively, where $\textbf{R}_m$ is the radius vector of the molecule with the number $m$ and ${\bf d}_m$ is the transition dipole moment of the molecule $m$.
The vector ${\bf d}_m$ is directed for each molecule along the $x$ axis (Fig.~\ref{setup}).

Following the experimental setups presented in~\cite{anderson2018two, vento2023measurement}, we consider the scattering of two consecutive quasi-monochromatic pulses on the ensemble of the molecules. 
The first pulse is a write pulse, and the second is a read pulse. We introduced a time delay $\tau$ between the read and the write pulses. 
The pulses propagate along the optical axis of the system $z$ as we show in Fig.~\ref{setup}. 
Since the optical system focuses the write and the read pulses on the sample, we represent their Fourier components as Gaussian beams. 
Parameter $z_0$ of the Gaussian beam characterizes the distance along $z$ axis at which the beam radius becomes $\sqrt{2}$ times greater than the radius of the beam waist $w_0$~\cite{hohenester2020nano, novotny2012principles}. 
For example, the He-Ne laser has $z_0\sim1$ mm when $w_0=20$ $\mu$m~\cite{saleh2019fundamentals}. We consider the case $l_{\rm sample} \ll z_0$, where $l_{\rm sample}$ is the length of the sample along $z$ axis.
In this case, the interaction constants of the pulses with the molecules are (Appendix~\ref{focal_field_derivation})
\begin{equation} \label{Rabi frequency}
\Omega_m^{\rm (write/read)}(t)=A^{\rm (write/read)}e^{-R_m'^2/2w_0^2}F^{\rm (write/read)}\left(t-\frac{\tilde n z_m}{c}\right)e^{-i\omega^{\rm (write/read)}(t-\tilde n z_m/c)},
\end{equation}
where $A^{\rm (write/read)}$ are the maximal amplitudes of the read and write pulses, $F^{\rm (write/read)}(t)$ are the temporal profile of these pulses normalized by the condition $\max\limits_t [F^{\rm (write/read)}(t)] = 1$, $\omega^{\rm (write/read)}$ are their central frequencies, ${{\bf R}'_m}$ is projection of ${{\bf R}_m}$ onto a plane $xy$ (Fig.~\ref{setup}), $z_m$ is the $z$ axis coordinate of the $m$th molecule, and $\tilde n$ is the refractive index of the media.

The operator of the dipole moment of the $m$th molecule is ${\bf d}_m \hat \sigma_m$~\cite{scully1997quantum,carmichael2013statistical}, thus $\hat \sigma_m$ determines the statistical properties of the scattered light.
Molecules interact with an environment which causes relaxation processes. 
One can elimanate environment degrees of freedom using Born-Markov approximation and obtain Heisenberg--Langevin equations~\cite{gardiner2004quantum}~(Appendix \ref{hl_equations}).
Heisenberg--Langevin equations allow us to find an approximate expression for the operator $\hat\sigma_m(t)$ in the Heisenberg representation in the case of a weak external field, i.e. the small parameter is $|A^{\rm (write/read)}/(\omega^{(\rm el)}_m - \omega^{\rm (write/read)})|\ll 1$. 
We consider nonresonant Raman scattering, $|\omega^{(\rm el)}_m - \omega^{\rm (write/read)}|\gg\gamma_{\perp}$, where $\gamma_{\perp}$ is the width of the electronic transition line of the molecules.
The resultant expression for $\hat\sigma_m(t)$ obtained in~\cite{shishkov2021enhancement, lisyansky2024quantum} is $\hat\sigma_m(t) \approx \hat \sigma_m^{\rm (R)}(t) + \hat \sigma_m^{\rm (St)}(t) + \hat \sigma_m^{\rm (aSt)}(t) + \hat \sigma_m^{\rm (R')}(t)$ contains the terms responsible for Rayleigh scattering $\hat \sigma_m^{\rm (R)}(t)$, Stokes scattering from the write pulse, $\hat \sigma_m^{\rm (St)}(t)$, and anti-Stokes scattering from the read pulse, $\hat \sigma_m^{\rm (aSt)}(t)$, and all the rest components of Raman scattering, $\hat \sigma_m^{\rm (R')}(t)$, where
\begin{equation} \label{sigma_St}
\hat \sigma_m^{\rm (St)}(t)
=
\sum\limits_{j=1}^{N^{\rm (vib)}_m}
\frac{\Lambda_{mj}\omega^{\rm (vib)}_{mj}}{\omega^{\rm (write)}-\omega^{\rm (vib)}_{mj}-\omega^{\rm (el)}_m}
\frac{\Omega^{\rm (write)}_m(t)}{\omega^{\rm (write)}-\omega^{\rm (el)}_m}\hat b^{\dag}_{mj}(t),
\end{equation}
\begin{equation} \label{sigma_aSt}
\hat \sigma_m^{\rm (aSt)}(t)
=
\sum\limits_{j=1}^{N^{\rm (vib)}_m}
\frac{\Lambda_{mj}\omega^{\rm (vib)}_{mj}}{\omega^{\rm (read)}+\omega^{\rm (vib)}_{mj}-\omega^{\rm (el)}_m}
\frac{\Omega^{\rm (read)}_m(t-\tau)}{\omega^{\rm (read)}-\omega^{\rm (el)}_m}\hat b_{mj}(t).
\end{equation}
Here $\hat b_{mj}(t)$ describes the vibrations of the nuclei of a molecule under the influence of the thermal and have the correlations $\langle \hat b_{mj}^\dag(t) \hat b_{m'j'}(t+\tau) \rangle = \delta_{j'j}\delta_{m'm} n^{\rm (th)}_{mj} e^{-i(\omega^{\rm (vib)}_{mj}-i\gamma^{\rm (vib)}_{mj})\tau}$ and $\langle \hat b_{mj}(t) \hat b_{m'j'}(t) \rangle = 0$, where $\gamma^{\rm (vib)}_{mj}$ is the decay rate of the vibrational amplitude, $n^{\rm (th)}_{mj} = (e^{\hbar\omega^{\rm (vib)}_{mj}/k_{\rm B}T} - 1)^{-1}$, $T$ is the ambient temperature, and $k_B$ is Boltzmann costant.
We note that the decay rate of the vibrational amplitude is primaraly determined by dephasing processes in liquids and by dissipation processes in gases~\cite{tokmakoff1995infrared, rector1997vibrational, velez2019preparation}.

In the setup shown in Fig.~\ref{setup}, the molecules are in the focus of the second lens such that the fiber collecting the light is in the focus plane of the third lens. 
We consider linearly polarized incident light with the polarization along $x$ axes~(Fig.~\ref{setup}). 
We describe light propagation in the optical system in paraxial approximation.
In this case, the scattered light preserves its polarization while entering and propagating through the fiber.
Thus, we can refer to the electric field as a scalar.
The operators of the Stokes and anti-Stokes signals at the point $\bf r$ at the plane of the fiber input interface are (Appendix~\ref{molecules_field_derivation})
\begin{equation} \label{E}
\hat E_{\rm in}^{\rm (St/aSt)} ({\bf r}, t)=D^{\rm (St/aSt)}\sum_{m=1}^{N^{\rm (mol)}}\hat \sigma_m^{\rm (St/aSt)}\left(t-t_0+\frac{\tilde n z_m}{c}\right)u^{\rm (St/aSt)}({\bf r}, {\bf R}_m),
\end{equation}
where $D^{\rm (St/aSt)}$ and $t_0$ are parameters on depending the optical setup~\cite{hohenester2020nano, novotny2012principles}, $u^{\rm (St/aSt)}({\bf r}, {\bf R}_m)$ are the spot profiles of the scattered light from $m$th molecule radiation at the input plane of the fiber (here, ${\bf r}$ is two-dimensional vector).

\section{Light propagation in a weakly guiding fiber} \label{sec:propagation}
A light propagating in a weakly guiding fiber is a superposition of linearly polarized modes~\cite{solimeno2012guiding}.
Since the electric field of the scattered light has linear polarization along the $x$ axis, this light excites only fiber modes with the same polarization.
Therefore, we consider only the modes of the fiber polarized along the $x$ axis (see Fig.~\ref{setup}).
We assume the knowledge of the parameters of all the propagating modes in the fiber for any given frequency $\omega$.
For the detailed analysis of the modes, including the expressions for field distribution of these modes, we refer to~\cite{solimeno2012guiding}.
We denote the number of propagating modes polarized along $x$ axis in the fiber at a frequency $\omega$ as $N^{\rm (modes)}_{{\rm LP}}(\omega)$, where the subscript ``LP'' stands for linearly polarized.
We also denote the electric field distribution in the cross-section of the fiber of mode $k$ at a frequency $\omega$ as $f_k({\bf r}, \omega)$ and the corresponding constant of propagation as $\beta_k(\omega)$.
The propagating modes are orthogonal, $\int f_k({\bf r}, \omega)f_{k'}({\bf r}, \omega)d^2{\bf r}=\delta_{kk'}$, where the integration is in the plane perpendicular to the fiber axis. Note that $f_k({\bf r}, \omega)$ are real-valued functions.

We assume that Stokes and anti-Stokes components of the scattered light are quasi-monocromatic with the corresponding central frequencies $\omega^{\rm (St)}$ and $\omega^{\rm (aSt)}$.
We assume that $|\omega^{\rm (aSt)}-\omega^{\rm (St)}| \ll \omega^{\rm (St/aSt)}$ and $N^{\rm (modes)}_{\rm LP}(\omega^{\rm (St)})=N^{\rm (modes)}_{\rm LP}(\omega^{\rm (aSt)})$. 
These condtitions imply that modes $f_k({\bf r},\omega^{\rm (St)})$ and $f_{k'}({\bf r},\omega^{\rm (aSt)})$ are almost orthogonal, $\int f_k({\bf r}, \omega^{\rm (St)})f_{k'}({\bf r}, \omega^{\rm (aSt)}) d^2{\bf r} \approx \delta_{kk'}$.
We denote $N^{\rm (modes)}_{\rm LP} = N^{\rm (modes)}_{\rm LP}(\omega^{\rm (St)}) = N^{\rm (modes)}_{\rm LP}(\omega^{\rm (aSt)})$, $f_k^{\rm (St)}({\bf r}) = f_k({\bf r},\omega^{\rm (St)})$, $f_k^{\rm (aSt)}({\bf r}) = f_k({\bf r},\omega^{\rm (aSt)})$.

We assume that the fiber is long enough so that only guided modes contribute to the output signal.
We neglect nonlinear effects and non-radiative losses in the fiber.
Projecting the field on these modes at the input interface of the fiber and taking into account the quasi-monochromaticity of the signals, we obtain the electric field of the Stokes and anti-Stokes output signals 
\begin{equation} \label{Output field}
\hat E^{\rm (St/aSt)}_{\rm out}({\bf r},t)
\approx A^{\rm (St/aSt)} 
\sum\limits_{k=1}^{N^{\rm (modes)}_{\rm LP}} f^{\rm (St/aSt)}_k({\bf r})
\int\limits_{-\infty}^{t} dt' 
G_k(t-t')
\hat E^{\rm (St/aSt)}_k(t'),
\end{equation}
where 
$
G_k(t-t')
=
(2\pi)^{-1}
\int_{-\infty}^{+\infty}  
e^{i \beta_k(\omega)L}e^{-i\omega (t-t')}
d\omega
$
and 
\begin{equation} \label{Stokes_mode}
\hat E^{\rm (St/aSt)}_k(t)
=
\int \hat E^{\rm (St/aSt)}_{\rm in}({\bf r},t) f^{\rm (St/aSt)}_k({\bf r}) d^2{\bf r} 
,
\end{equation}
where $L$ is the fiber length, vector $\bf r$ is at the output interface plane of the fiber, and the integration over $\bf r$ is going in the plane of input interface on the fiber.
The coefficients $A^{\rm (St/aSt)}$ correspond to the difference between the incident field and transmitted field due to the reflection on the input and output interfaces.  
The upper limit of time integration in Eq.~(\ref{Output field}) equals $t$ that agrees with the causality principle.


\section{Coherent properties of the Raman light}

\subsection{Cross-correlation function}

We analyze the second-order cross-correlation functions between the Stokes signal from the scattering of the write pulse and the anti-Stokes signal from the scattering of the read pulse.
We assume that the Stokes and the anti-Stokes signals are spectrally separated.
Typically, the detector response time exceeds the duration of the Stokes and anti-Stokes signals.
For instance, an EMCCD camera has a few milliseconds time response~\cite{edgar2012imaging}, while the Stokes and anti-Stokes signals are around $500$~fs long~\cite{vento2023measurement, anderson2018two}.
Thus, we obtain the second-order cross-correlation function
\begin{multline} \label{g2} 
\tilde g^{(2)}_{{\rm St}, {\rm aSt}}(\tau)\\
=
\frac{
\int\limits_{-\infty}^{+\infty} dt_1\int\limits_{-\infty}^{+\infty} dt_2\int d^2{\bf r}_1\int d^2{\bf r}_2 
\langle 
\mathcal{T}_\rightarrow 
\{  
\hat E^{({\rm St}) \dag}_{\rm out} ({\bf r}_1, t_1) \hat E^{({\rm aSt})\dag}_{\rm out} ({\bf r}_2,t_2) 
\}
\mathcal{T}_\leftarrow  
\{  
\hat E^{({\rm aSt})}_{\rm out}({\bf r}_2, t_2) \hat E^{({\rm St})}_{\rm out} ({\bf r}_1, t_1)  
\}
\rangle
}{
\int\limits_{-\infty}^{+\infty} dt
\int d^2{\bf r} 
\langle \hat E^{\rm (St)\dag}_{\rm out} ({\bf r}, t) \hat E^{\rm (St)}_{\rm out}({\bf r},t) \} \rangle 
\int\limits_{-\infty}^{+\infty} dt
\int d^2{\bf r} 
\langle \hat E^{\rm (aSt)\dag}_{\rm out} ({\bf r}, t) \hat E^{\rm (aSt)}_{\rm out}({\bf r},t) \} \rangle 
}.
\end{multline}
where $\mathcal{T}_\rightarrow$ and $\mathcal{T}_\leftarrow$ are the time ordering over the source operators~\cite{knoll1986quantum, vogel2006quantum}.
In our case, the source operators are $\hat \sigma_m^{\rm (St/aSt)}(t)$.
Using Eq.~(\ref{Output field}), we obtain (Appendix~\ref{pm_derivation})
\begin{multline} \label{physical_meaning}
\tilde g^{(2)}_{\rm St,aSt}(\tau)=1 
+\frac{
\sum\limits_{k_1=1}^{N^{\rm (modes)}_{\rm LP}}
\sum\limits_{k_2=1}^{N^{\rm (modes)}_{\rm LP}}
\int\limits_{-\infty}^{+\infty}dt_2
\int\limits_{-\infty}^{t_2}dt_1 
\langle \hat E^{\rm (St)\dag}_{k_1}(t_1) \hat E^{\rm (aSt)\dag}_{k_2}(t_2) \rangle 
\langle \hat E^{\rm (aSt)}_{k_2}(t_2) \hat E^{\rm (St)}_{k_1}(t_1) \rangle
}{
\sum\limits_{k_1=1}^{N^{\rm (modes)}_{\rm LP}}
\sum\limits_{k_2=1}^{N^{\rm (modes)}_{\rm LP}}
\int\limits_{-\infty}^{+\infty} dt_1
\int\limits_{-\infty}^{+\infty} dt_2
\langle \hat E^{\rm (St)\dag}_{k_1}(t_1) \hat E^{\rm (St)}_{k_1}(t_1) \rangle 
\langle \hat E^{\rm (aSt)\dag}_{k_2}(t_2) \hat E^{\rm (aSt)}_{k_2}(t_2) \rangle 
}
\\
+
\frac{
\sum\limits_{k_1=1}^{N^{\rm (modes)}_{\rm LP}}
\sum\limits_{k_2=1}^{N^{\rm (modes)}_{\rm LP}}
\int\limits_{-\infty}^{+\infty}dt_1
\int\limits_{-\infty}^{t_1}dt_2 
\langle \hat E^{\rm (aSt)\dag}_{k_2}(t_2) \hat E^{\rm (St)\dag}_{k_1}(t_1) \rangle 
\langle \hat E^{\rm (St)}_{k_1}(t_1) \hat E^{\rm (aSt)}_{k_2}(t_2) \rangle
}{
\sum\limits_{k_1=1}^{N^{\rm (modes)}_{\rm LP}}
\sum\limits_{k_2=1}^{N^{\rm (modes)}_{\rm LP}}
\int\limits_{-\infty}^{+\infty}dt_1
\int\limits_{-\infty}^{+\infty}dt_2
\langle \hat E^{\rm (St)\dag}_{k_1}(t_1) \hat E^{\rm (St)}_{k_1}(t_1) \rangle 
\langle \hat E^{\rm (aSt)\dag}_{k_2}(t_2) \hat E^{\rm (aSt)}_{k_2}(t_2) \rangle 
}.
\end{multline}
When deriving~(\ref{physical_meaning}), we used that Stokes and anti-Stokes signals are hosted by thermal oscillations of molecules nuclei taking into account Eq.~(\ref{sigma_St}) and Eq.~(\ref{sigma_aSt}).

Knowledge of the paired correlations of operators~(\ref{Stokes_mode}), allows us to express $\tilde g^{(2)}_{\rm St,aSt}(\rm \tau)$ through the parameters of the system. 
In Appendix~\ref{corr_derivation}, we derive paired correlations that allow us to express $\tilde g^{(2)}_{\rm St,aSt}(\rm \tau)$.
Thus, we obtain 
\begin{equation} \label{g2ideal}
\tilde g^{(2)}_{\rm St, aSt}(\tau)
=
1
+
\frac{1}{B^{\rm ({\rm St, aSt})}_{\rm coh}B^{\rm (St,aSt)}_{\rm fib}}
\frac{1 + n^{(\rm th)}_{\rm vib} }{ n^{(\rm th)}_{\rm vib} }
\\
+
\frac{1}{B^{\rm ({\rm aSt, St})}_{\rm coh}B^{\rm (aSt,St)}_{\rm fib}}
\frac{n^{(\rm th)}_{\rm vib} }{ 1 + n^{(\rm th)}_{\rm vib} },
\end{equation}
where
\begin{equation} \label{coh_factor}
    B_{\rm coh}^{(\alpha,\beta)}=
\frac{
\tau^{\rm (write)}
\tau^{\rm (read)}
\left[
\sum\limits_{m=1}^{N^{\rm (mol)}}
\sum\limits_{j=1}^{N^{\rm (vib)}_{m}}
\Lambda_{mj}^2
\right]^2
}{
\sum\limits_{m_1=1}^{N^{\rm (mol)}}
\sum\limits_{j_1=1}^{N^{\rm (vib)}_{m_1}}
\sum\limits_{m_2=1}^{N^{\rm (mol)}}
\sum\limits_{j_2=1}^{N^{\rm (vib)}_{m_2}}
I_{m_1j_1m_2j_2}^{\rm (\alpha,\beta)}
\Lambda_{m_1j_1}^2 \Lambda_{m_2j_2}^2
},
\end{equation}
\begin{equation} \label{B_fib}
B^{\rm (\alpha,\beta)}_{\rm fib}=\frac{\sum_{k_1=1}^{N^{\rm (modes)}_{\rm LP}}\sum_{k_2=1}^{N^{\rm (modes)}_{\rm LP}}(F_{k_1}^{\rm (\alpha)},F_{k_1}^{\rm (\alpha)})
(F_{k_2}^{\rm (\beta)},F_{k_2}^{\rm (\beta)})}
{\sum_{k_1=1}^{N^{\rm (modes)}_{\rm LP}}\sum_{k_2=1}^{N^{\rm (modes)}_{\rm LP}}|(F_{k_1}^{\rm (\alpha)*},F_{k_2}^{\rm (\beta)})|^2}.
\end{equation}
Here $\alpha, \beta\in \left\{ {\rm St}, {\rm aSt} \right\}$. 
The dependece of $\tilde g^{(2)}_{\rm St, aSt}(\tau)$ on $\tau$ lies in coefficients $B_{\rm coh}^{(\alpha,\beta)}$ which depend on $I_{m_1j_1m_2j_2}^{\rm (\alpha,\beta)}$.
We give the expressions for coefficients $I_{m_1j_1m_2j_2}^{\rm (\alpha,\beta)}$ in Appendix~\ref{I_coeff}.
We also use a notation $F^{\rm (St/aSt)}_k({\bf R}_m)=\int d^2{\bf r}f^{\rm (St/aSt)}_k({\bf r})u^{\rm (St/aSt)}({\bf r}, {\bf R}_m)$ and define the scalar product as $(F_1,F_2)=\int d^3{\bf R}_m  N({\bf R}_m)F^{*}_1({\bf R}_m)F_2({\bf R}_m)\exp\left(-R_m'^2/w_0^2\right)$ where $N({\bf R}_m)$ is the concentration of the molecules and the integration goes along the volume of the sample. 
Here, we used that the thermal occupations $n^{(\rm th)}_{mj}$ are approximately equal to $n^{(\rm th)}_{\rm vib}$ and we denoted the duration of the read and write pulses $\tau^{\rm (write/read)} = \int_{-\infty}^{+\infty} [ F^{\rm (write/read)}(t) ]^2 dt$ and the temporal overlaps of the scattered lights. 
One can show that in a case $\tau^{\rm (write)}=\tau^{\rm (read)}$ parameter $B^{\rm (\alpha,\beta)}_{\rm coh}$ decreases when pulses duration decreases.
Therefore, one need to use pulses with small durations to observe more high correlations.  
Note that the value $B_{{\rm{fib}}}^{(\alpha,\beta)}$ are the fiber-induced factors that suppress Stokes--anti-Stokes correlations, and $B_{\rm coh}^{(\alpha,\beta)}$ is the suppression factor of Stokes--anti-Stokes correlations due to loss of the coherence of the molecular vibrations between arrival of read and write pulses.

Both the numerator and the denominator of Eq.~(\ref{coh_factor}) are proportional to $[N^{\rm (mol)}]^2$, thus the second and the third terms in Eq.~(\ref{g2ideal}) do not depend on $[N^{\rm (mol)}]$.
This implies that the Stokes--anti-Stokes correlations do not vanish even when the number of molecules is macroscopic.
These results agree with our previous theoretical results~\cite{panyukov2022heralded} and are consistent with the experimental data~\cite{kasperczyk2016temporal}.
In our calculations we use an assumption $N^{\rm (mol)} \gg 1$~(\ref{N_approx})
In this limit correlations do not depend on $N^{\rm (mol)}$. 
However, our theory allows to explicitly take into account molecular correlations through the terms .

To illustrate the suppression of the second-order cross-correlation function in multi-mode fiber, we consider a limiting case when (1) geometric optics is applicable, $u^{\rm (St/aSt)}({\bf r},{\bf R}_m)\approx \delta({\bf r}+\tilde n f' {\bf R}'_m/f)$, and (2) the image of the illuminated part of the sample covers the whole fiber core, ${\min}\{w_0, R_{\rm sample}\} \gg a f/\tilde n f'$, where $R_{\rm sample}$ is the sample size and $a$ is the radius of the fiber core (see Fig.~\ref{setup}),
(3) the concentration $N({\bf R}_m)$ does not depend on ${\bf R}_m$.
In this case, $\langle \hat E^{\rm (St)}_{k_1}(t_1) \hat E^{\rm (aSt)}_{k_2}(t_2) \rangle \propto \delta_{k_1k_2}$, implying that the quantum amplitudes corresponding to the excitation of different fiber modes are uncorrelated.
It follows that  $B^{\rm (St,aSt)}_{\rm fib}\approx N^{\rm (modes)}_{\rm LP}$, therefore the non-classical contribution to the correlations $\tilde g^{(2)}_{\rm St, aSt}(\tau)$ is inversely proportional to the number of propagating modes $N^{\rm (modes)}_{\rm LP}$.

In general case Cauchy–Schwarz inequality implies 
\begin{equation}
|(F_{k_1}^{\rm (St)*},F_{k_2}^{\rm (aSt)})|^2 \leqslant (F_{k_1}^{\rm (St)},F_{k_1}^{\rm (St)}) (F_{k_2}^{\rm (aSt)},F_{k_2}^{\rm (aSt)}),
\end{equation}
therefore $B^{\rm (St,aSt)}_{\rm fib} \geqslant  1$ and it increases with the number of propagating fiber modes.
Thus, $\tilde g^{(2)}_{\rm St, aSt}(\tau)$ decreases as the number of modes increases, which agrees with the recent experiments~\cite{vento2023measurement}.

Any additional source of uncorrelated photons may suppress the second-order cross-correlation function.
The background-adjusted second-order cross-correlation function of the Stokes and anti-Stokes light, $g^{(2)}_{\rm St, aSt}(\tau)$, depends on the signal-to-noise ratio for Stokes signal, ${\rm SNR^{(St)}}$, and anti-Stokes signal, ${\rm SNR^{(aSt)}}$ $g^{(2)}_{\rm St, aSt}(\tau) = 1 + \left(\tilde g^{(2)}_{\rm St, aSt}(\tau)-1\right)/B^{\rm (St,aSt)}_{\rm bg}$ where $B^{\rm (St,aSt)}_{\rm bg}=(1+{1}/{\rm SNR^{(St)}})(1+{1}/{\rm SNR^{(aSt)}})$~\cite{panyukov2022heralded}.
Since the intensity of the anti-Stokes signal is smaller than the intensity of the Stokes signal, the background light in the anti-Stokes spectral region is more crucial for the second-order cross-correlation function. 
The combined effect of loss of the temporal coherence of molecular vibrations, the light propagation through the fiber, and the background light suppresses the second-order cross-correlation function by the factor $B^{\rm (St,aSt)}_{\rm coh}B^{\rm (St,aSt)}_{\rm fib}B^{\rm (St,aSt)}_{\rm bg}$.
In the absence of these three factors $B^{\rm (St,aSt)}_{\rm coh}B^{\rm (St,aSt)}_{\rm fib}B^{\rm (St,aSt)}_{\rm bg}=1$ and $g^{(2)}_{\rm St, aSt}(0) \approx 1/n^{\rm (vib)}_{\rm th}$, reaching its quantum limit.

\subsection{Second-order autocorrelation function}

Our theory also applies to the second-order autocorrelation function of the Stokes and anti-Stokes signals.  
Similarly to Eq.~(\ref{g2}), the second-order coherence function of the Stokes light is
\begin{multline} \label{g2_new} 
\tilde g^{(2)}_{\alpha, \alpha}(\tau)\\
=
\frac
{
\int\limits_{-\infty}^{+\infty} dt_1\int\limits_{-\infty}^{+\infty} dt_2\int d^2{\bf r}_1\int d^2{\bf r}_2 
\langle 
\mathcal{T}_\rightarrow 
\{  
\hat E^{(\alpha)\dag}_{\rm out} ({\bf r}_1, t_1) \hat E^{(\alpha)\dag}_{\rm out} ({\bf r}_2,t_2) 
\}
\mathcal{T}_\leftarrow  
\{  
\hat E^{(\alpha)}_{\rm out}({\bf r}_2, t_2) \hat E^{(\alpha)}_{\rm out} ({\bf r}_1, t_1)  
\}
\rangle 
}
{
\int\limits_{-\infty}^{+\infty} dt
\int d^2{\bf r} 
\langle \hat E^{(\alpha)\dag}_{\rm out} ({\bf r}, t) \hat E^{(\alpha)}_{\rm out}({\bf r},t) \} \rangle 
\int\limits_{-\infty}^{+\infty} dt
\int d^2{\bf r} 
\langle \hat E^{(\alpha)\dag}_{\rm out} ({\bf r}, t) \hat E^{(\alpha)}_{\rm out}({\bf r},t) \} \rangle 
},
\end{multline}
where $\alpha \in \left\{ {\rm St}, {\rm aSt} \right\}$. 
From Eq.~(\ref{Output field}), we obtain the expression for $\tilde g^{(2)}_{\rm \alpha, \alpha}(\tau)$
\begin{multline} \label{physical_meaning_new}
\tilde g^{(2)}_{\alpha, \alpha}(\tau)=1 
+\frac{
\sum\limits_{k_1=1}^{N^{\rm (modes)}_{\rm LP}}
\sum\limits_{k_2=1}^{N^{\rm (modes)}_{\rm LP}}
\int\limits_{-\infty}^{+\infty}dt_2
\int\limits_{-\infty}^{t_2}dt_1 
\langle \hat E^{\rm (\alpha)\dag}_{k_1}(t_1) \hat E^{\rm (\alpha)}_{k_2}(t_2) \rangle 
\langle \hat E^{\rm (\alpha)\dag}_{k_2}(t_2) \hat E^{\rm (\alpha)}_{k_1}(t_1) \rangle
}{
\sum\limits_{k_1=1}^{N^{\rm (modes)}_{\rm LP}}
\sum\limits_{k_2=1}^{N^{\rm (modes)}_{\rm LP}}
\int\limits_{-\infty}^{+\infty} dt_1
\int\limits_{-\infty}^{+\infty} dt_2
\langle \hat E^{\rm (\alpha)\dag}_{k_1}(t_1) \hat E^{\rm (\alpha)}_{k_1}(t_1) \rangle 
\langle \hat E^{\rm (\alpha)\dag}_{k_2}(t_2) \hat E^{\rm (\alpha)}_{k_2}(t_2) \rangle 
}
\\
+
\frac{
\sum\limits_{k_1=1}^{N^{\rm (modes)}_{\rm LP}}
\sum\limits_{k_2=1}^{N^{\rm (modes)}_{\rm LP}}
\int\limits_{-\infty}^{+\infty}dt_1
\int\limits_{-\infty}^{t_1}dt_2 
\langle \hat E^{\rm (\alpha)\dag}_{k_2}(t_2) \hat E^{\rm (\alpha)}_{k_1}(t_1) \rangle 
\langle \hat E^{\rm (\alpha)\dag}_{k_1}(t_1) \hat E^{\rm (\alpha)}_{k_2}(t_2) \rangle
}{
\sum\limits_{k_1=1}^{N^{\rm (modes)}_{\rm LP}}
\sum\limits_{k_2=1}^{N^{\rm (modes)}_{\rm LP}}
\int\limits_{-\infty}^{+\infty}dt_1
\int\limits_{-\infty}^{+\infty}dt_2
\langle \hat E^{\rm (\alpha)\dag}_{k_1}(t_1) \hat E^{\rm (\alpha)}_{k_1}(t_1) \rangle 
\langle \hat E^{\rm (\alpha)\dag}_{k_2}(t_2) \hat E^{\rm (\alpha)}_{k_2}(t_2) \rangle 
}.
\end{multline}
We express $\tilde g^{(2)}_{\alpha, \alpha}(\tau)$ through the parameters of the system
\begin{equation} \label{g2ideal_new}
\tilde g^{(2)}_{\alpha, \alpha}(\tau)
=
1
+
\frac{1}{B_{\rm coh}^{(\alpha,\alpha)} B^{\rm (\alpha, \alpha)}_{\rm fib}},
\end{equation}
where ${B_{\rm coh}^{(\alpha,\alpha)}}$ and ${B^{\rm (\alpha,\alpha)}_{\rm fib}}$ are defined by Eq.~(\ref{coh_factor}) and Eq.~(\ref{B_fib}) respectively.
We can take into account the background light in the same way we did it for the second-order cross-correlation function, obtaining $g^{(2)}_{\rm St, St}(0) = 1 + \left(\tilde g^{(2)}_{\rm St, St}(0)-1\right)/B^{\rm (St,St)}_{\rm bg}$ and $g^{(2)}_{\rm aSt,aSt}(0) = 1 + \left(\tilde g^{(2)}_{\rm aSt, aSt}(0)-1\right)/B^{\rm (aSt, aSt)}_{\rm bg}$.
In the limiting case considered above, $u^{\rm (St/aSt)}({\bf r},{\bf R}_m)\approx \delta({\bf r}+\tilde n f' {\bf R}'_m/f)$ and ${\rm min}\{w_0, R_{\rm sample}\} \gg af/\tilde n f'$, it is easy to show that $B^{\rm (St,St)}_{\rm fib} \approx B^{\rm (aSt,aSt)}_{\rm fib}\approx N^{\rm (modes)}_{\rm LP}$.
As the number of modes $N^{\rm (modes)}_{\rm LP}$ increases, $g^{(2)}_{\rm St, St}(0)$ and $g^{(2)}_{\rm aSt, aSt}(0)$ tend to 1, while the non-classical contribution tends to zero inversely proportional to $N^{\rm (modes)}_{\rm LP}$. 

\subsection{Simple explanation of correlations suppression in multi-mode optical fiber.}

\begin{figure}
\includegraphics[width=1\linewidth]{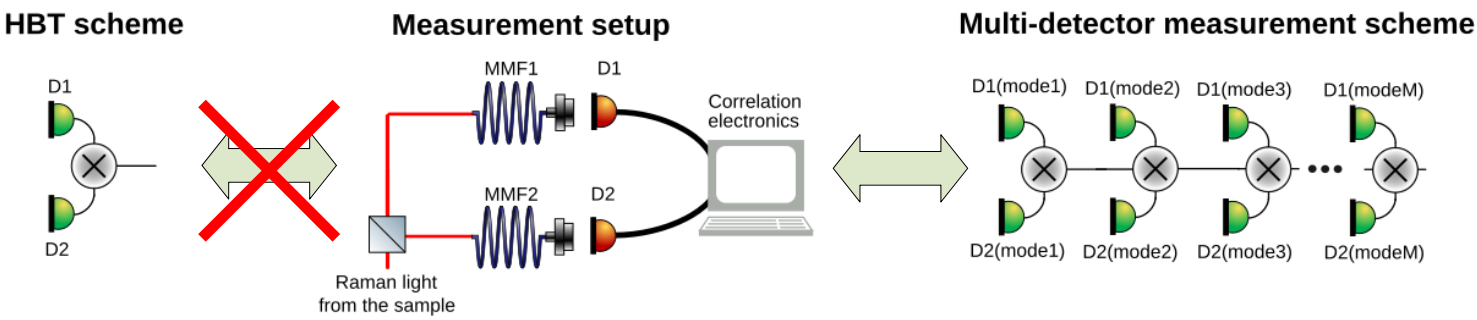}
\caption{
The equivalent representation of ``HBT scheme'' with multi-mode fibers illustrating what is actually measured in the ``HBT scheme'' with multi-mode fibers.
Central figure is the measurement scheme used in many experiments (for example~\cite{anderson2018two, vento2023measurement}) on Stokes--anti-Stokes correlations.
MMF1 and MMF2 are multi-mode fibers, D1 and D2 are detectors connected to these fibers.
Figure on the left is a standard HBT scheme with two detectors.
Figure on the right is multi-detector measurement scheme equivalent to the experimental setup.
D1(mode$N$) and D2(mode$N$) are the set of detectors (one for each fiber mode) equivalent to the D1 and D2 in the central figure (real experiment setup).
} \label{Equivalent_measurement_scheme}
\end{figure}

One of the most prominent consequence of our theoryis that the ``HBT scheme'' with multi-mode fibers (MMF) is not equivalent to the true free-optics HBT scheme (Fig.~\ref{Equivalent_measurement_scheme}).
Thus, ``HBT scheme'' with multi-mode fibers does not measure the true second-order autocorrelation or second-order cross-correlation functions.
Here, we discuss the origins of such non-equivalence and what is actually measured in the ``HBT scheme'' with MMF, presenting the equivalence scheme.

The reason why ``HBT scheme'' with MMFs does not measure true second-order autocorrelation or second-order cross-correlation functions is the connection between the fiber and the detector.
The layout of typical connection between MMF and detector can be found for example in~\cite{idq100}.
The fiber is mounted in a certain position with fixed distance from the detector with no additional optical elements between end of the fiber and detector.
This distance is chosen in such a way that the emission pattern from the fiber in real space maximally overlaps the detector.
The electric fields of different modes of the fiber preserves orthogonality while propagating freely from the fiber to the detector.
Due to the orthogonality of the light emitted from different modes at detector, the detector registers photons from each fiber mode independently.
Thus, the equivalent scheme of ``HBT scheme'' with multi-mode fibers is the set of independent detectors for each of the fiber modes shown in Fig.~\ref{Equivalent_measurement_scheme}.
We call this equivalent scheme ``Multi-detector measurement scheme''.
In this scheme the electromagnetic field each MMF is detected with its own detector, but the coincidence scheme does not distinguish between the different modes, meaning that, for instance, the clicks in D1(mode 3) and D2(mode 1) are the coincidence event in this scheme.
Such events suppress the overall measured correlations both for cross-correlations and autocorrelation measurements.
Thus, this equivalent scheme explains the suppression of the correlations when they are measured with multi-mode optical fibers.

\section{Numerical simulation}
Similarly to the experimental setup~\cite{vento2023measurement}, we consider the liquid, consisting of two types of molecules.
The molecules of the first type have three vibrational modes, while the molecules of the second type have one vibrational mode~\cite{vento2023measurement}.
The thermal occupation of the vibrational modes of the molecules approximately equal $n^{\rm (th)}_{\rm vib}=0.04$.
The write and read pulses are Gaussian with temporal profiles $F^{\rm (write)}(t)=F^{\rm (read)}(t)=\exp(-t^2/2\sigma^2)$, where the pulse durations are $\tau^{\rm (write)}/\sqrt{\pi}=\tau^{\rm (read)}/\sqrt{\pi}=\sigma=0.2$~ps.

\begin{figure}
	\includegraphics[width=1\linewidth]{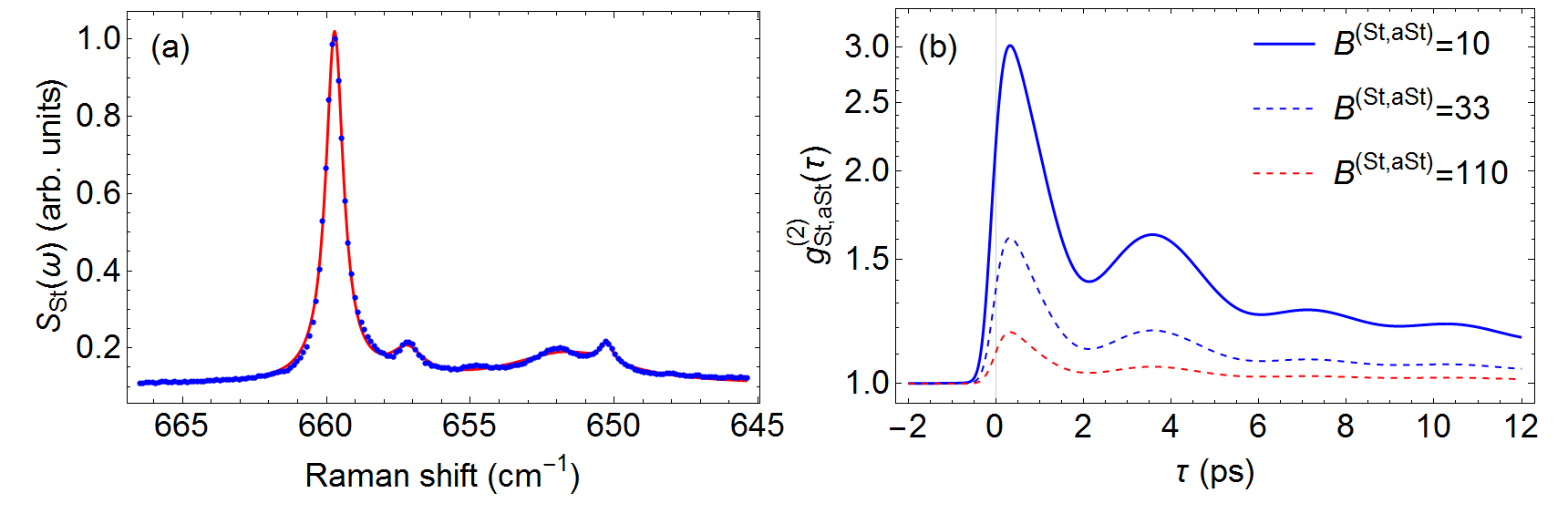}
	\caption{(a) The Stokes part of the CW spectrum, (blue dotted) experimental dara from Ref. 2, (red dashed) our approximation $S_{\rm St}(\omega)$ (b) $g^{(2)}_{\rm St, aSt}(\tau)$ at different values of $B^{\rm (St,aSt)}$.
} \label{pictures}
\end{figure}

To calculate $g^{(2)}_{\rm St, aSt}(\tau)$, we also need the vibrational frequencies,  the coherence times and the ratios between $N_m\Lambda_{mj}^2$, where $N_m$ is the fraction of molecules of the $m$th type. 
We extract these parameters from the spectrum of spontaneous Raman scattering on our ensemble of molecules in the CW regime measured in~\cite{vento2023measurement}.
Fig.~\ref{pictures}a shows the Stokes part of the CW spectrum.
This spectrum contains four peaks, which we approximate within the framework of our model by~\cite{lisyansky2024quantum}
\begin{equation} \label{lorentz_spectrum}
S_{\rm St}(\omega)=S_0+\sum\limits_{m=1}^{N_{\rm mol}}\sum\limits_{j=1}^{N_m^{\rm (vib)}}\frac{1}{\pi}\frac{N_{m}\Lambda_{mj}^2\gamma_{mj}^{\rm (vib)}\omega_{mj}^{{\rm (vib)}2}}{\gamma_{mj}^{{\rm (vib)}2}+(\omega-\omega^{\rm (CW)}+\omega_{mj}^{\rm (vib)})^2},
\end{equation}
where $\omega^{\rm (CW)}$ is the frequency of the external field, $S_0$ is the frequency-independent contribution of background radiation.
Thus, we extracted $N_1\Lambda_{11}^2:N_1\Lambda_{12}^2:N_1\Lambda_{13}^2:N_2\Lambda_{21}^2=0.57:0.08:0.33:0.02$,
$\omega_{11}^{\rm (vib)}/2 \pi=19.78$~THz, $\omega_{12}^{\rm (vib)}/2 \pi=19.70$~THz, $\omega_{13}^{\rm (vib)}/2 \pi=19.54$~THz, $\omega_{21}^{\rm (vib)}/2 \pi=19.49$~THz,
$1/\gamma_{11}^{\rm (vib)}=14.21$~ps, $1/\gamma_{12}^{\rm (vib)}=7.73$~ps, $1/\gamma_{13}^{\rm (vib)}=2.22$~ps, $1/\gamma_{21}^{\rm (vib)}=21.97$~ps.

Using the data above, we obtain $g^{(2)}_{\rm St, aSt}(\tau)$, where we choose different values of the parameter $B^{\rm (St,aSt)}=B_{\rm fib}^{\rm (St,aSt)}B_{\rm bg}^{\rm (St,aSt)}$ ({Fig.~\ref{pictures}b}), which represents the suppression of the correlations as the number of modes increases and due to background radiation.
Our model reproduces the oscillations of $\tilde g^{(2)}_{\rm St, aSt}(\tau)$ with delay time between write and read pulses, $\tau$ (Fig.~\ref{pictures}b).
The frequencies of the oscillations are determined by the differences between the frequencies of molecular vibrations.
The oscillations indicate the macroscopic coherence between the vibrations hosted by different molecules.
The second-order cross-correlation function, $g^{(2)}_{\rm St, aSt}(\tau)$, calculated within the framework of our model demonstrates qualitative agreement with the experiment~\cite{vento2023measurement}. 

We also calculate the second-order autocorrelation functions for the Stokes and anti-Stokes light $g^{(2)}_{\rm St, St}(0) \approx 1+0.96/B^{\rm (St,St)}$ and $g^{(2)}_{\rm aSt, aSt}(0) \approx 1+0.96/B^{\rm (aSt,aSt)}$, where parameters $B^{\rm (St,St)}$ and $B^{\rm (aSt,aSt)}$ likewise represent the suppression of the correlations as the number of modes increases and due to background radiation.
These second-order autocorrelation functions evidence the incoherent properties of the Stokes and anti-Stokes light inherited from the thermal fluctuations of the molecular vibrations.

\section{CONCLUSION}

In this paper, we developed a theory for the cross-correlations and autocorrelations of Stokes and anti-Stokes spectral components produced in the scattering of two successive pulses (write pulse and read pulse) on an ensemble of molecules.
We focused on two particular effects: (1) oscillations of the second-order cross-correlation function of the light in the case of the molecules with multiple vibrational states and (2) the influence of optical fiber through which the Raman light propagates to the detectors on the statistical properties of this light.
We applied the microscopic theory of spontaneous Raman scattering based on the Heisenberg--Langevin equations~\cite{lisyansky2024quantum} to an ensemble of non-interacting molecules illuminated by two consecutive pulses (write pulse and read pulse).
We found the electric field operators of the Stokes light produced by the write pulse and the anti-Stokes light produced by the read pulse at the output interface of the optical waveguide.
With these operators, we obtained the cross-correlation function $g^{(2)}_{\rm St, aSt}(\tau)$ between these Stokes and anti-Stokes light as well as the corresponding second-order autocorrelation functions.

In the case the molecules hosting the vibrations with different frequencies, $g^{(2)}_{\rm St, aSt}(\tau)$ may experience beats as the function of time delay between the write and read pulses, $\tau$, with the frequencies determined by the difference in the vibrational frequencies.
Importantly, our theory applies even to the samples with several types of molecules having different vibrational modes.
The obtained results quantitatively agrees with the recent experiment~\cite{vento2023measurement}.

We showed that an increase in the number of fiber modes leads to the suppression of $g^{(2)}_{\rm St, aSt}(\tau)$, $g^{(2)}_{\rm St, St}(0)$, and $g^{(2)}_{\rm aSt, aSt}(0)$, which underlines the influence of the measurement setup on the Raman correlations of the light.
This result also quantitatively agrees with the recent experiments~\cite{vento2023measurement}.
We also showed that the background radiation leads to suppression of the correlations proportional to the background noise intensity.
The overall suppression factor includes the suppression factors due to multiple fiber modes and background radiation.

Note that the oscillatory behavior of $g^{(2)}_{ {\rm{St,aSt}}}$ can be reproduced without explicit consideration of molecular polarization degrees of freedom, as in~\cite{vento2023measurement}.
However, such consideration does not describe the dependence of macroscopic coherence on molecular positions as well as decreasing of the second-order cross-correlation function when signal propagates through multi-mode optical fiber.
Our theory is an expansion of the theory from~\cite{vento2023measurement} and explicitly consider a hamiltonian of each molecule interacting with the external field.
We describe it's electronic and vibrational subsystems and their interaction via Fröhlich term. 
Raman scattering is a direct consequence of equations in our theory, we don't need to introduce annihilation operators of Stokes and anti-Stokes fields in the hamiltonian. 
In our theory we obtain an electric field of the molecules explicitly from the solution of Heisenberg-Langevin equations via perturbation theory and this solution up to the second order contains terms responsible for Rayleigh, Stokes, and anti-Stokes scattering.

\begin{acknowledgments}
E.S.A. and I.V.P.  thank the Foundation for the Advancement of Theoretical Physics and Mathematics ``Basis''.
\end{acknowledgments}

\appendix 

\section{Interaction of the molecules with the write and read pulses} \label{focal_field_derivation}

In this Appendix, we derive the Rabi frequency~(\ref{Rabi frequency}).

First, we consider the write pulse.
The Fourier component with the frequency $\omega$ at the point of $m$th molecule, ${\bf R}_m$, is a Gaussian beam because of the lens and has the form
\begin{multline}
E_{\omega}^{\rm (write)}({\bf R}_m)=E^{\rm (write)}_{0\omega}\frac{w_0}{w(z_m)}\exp\left(-\frac{R_m'^2}{2w^2(z_m)}\right) \\
\exp\left( i \tilde n k z_m - i\arctan\frac{z_m}{z_0}+i\frac{\tilde n kR_m'^2}{2R(z_m)}\right),
\end{multline}
where $w_0$ is the narrowest waist of the beam, $z_0=\tilde n k w_0^2/2$, $w(z_m)=w_0\sqrt{1+z_m^2/z_0^2}$, $R(z_m)=z_m+z_0^2/z_m$, and $k=\omega/c$. 
Similarly to the main text, we assume that $l_{\rm sample} \ll z_0$, in this case, $E_{\omega}^{\rm (write)}({\bf R}_m)\approx E^{\rm (write)}_{0\omega}\exp\left(-R_m'^2/2w_0^2\right)\exp\left( i\tilde n kz_m \right)$. 
In the time domain the electric field of the write pulse, $E^{\rm (write)}({\bf R}_m,t)=\int E^{\rm (write)}_{\omega}({\bf R}_m)e^{-i\omega t}d\omega/2\pi$, takes the form $E^{\rm (write)}({\bf R}_m,t)=E^{\rm (write)}_0(t-\tilde n z_m/c)\exp\left(-R'^2_m/2w_0^2\right)$. 
Thus, we obtain the Rabi frequency of the write pulse 
\begin{equation}
\Omega_m^{\rm (write)}(t)=A^{\rm (write)}e^{-R_m'^2/2w_0^2}F^{\rm (write)}\left(t-\frac{\tilde n z_m}{c}\right)e^{-i\omega^{\rm (write)}(t-\tilde n z_m/c)},
\end{equation}
where $F^{\rm (write)}(t)$ is the dimensionless temporal envelop of the write pulse. 

The derivation of the Rabi frequency for the read pulse is similar.

\section{Heisenberg-Langevin equations}
\label{hl_equations}

Hamiltonian of the system interacting with the environment
\begin{equation}
	\hat H^{\rm (total)}=\hat H + \hat H^{\rm (env)} +\hat H^{\rm (int)}.
\end{equation}

We model  the environment as an ensemble of harmonic oscillators. Hamiltonian of the environment is

\begin{multline}
	\hat H^{\rm (env)}
	=\sum _{m=1}^{N^{\rm (mol)}} \sum_k \hbar \omega_{mk}^{(\rm el-diss)}\hat a_{mk}^\dag \hat a_{mk}
	+\sum _{m=1}^{N^{\rm (mol)}} \sum _{j=1}^{N_m^{\rm (vib)}} \sum_k \hbar \omega_{mjk}^{(\rm vib-diss)} \hat r_{mjk}^\dag \hat r_{mjk} \\
	+ \sum _{m=1}^{N^{\rm (mol)}} \sum _{j=1}^{N_m^{\rm (vib)}} \sum_k \hbar \omega_{mjk}^{(\rm vib-deph)} \hat c_{mjk}^\dag \hat c_{mjk}.
\end{multline}

The first term is responsible for the relaxation of the electronic subsystems of the molecules, the second term corresponds to the relaxation of the vibrational subsystems, and the third term describes the dephasing of the vibrational subsystems.

Interaction hamiltonian has the form
\begin{multline}
	\hat H^{\rm (int)}
	=\sum _{m=1}^{N^{\rm (mol)}} \sum_k \hbar \Omega_{mk}^{(\rm el-diss)}(\hat a_{mk}^\dag \hat \sigma 
	+ \hat a_{mk} \hat \sigma^\dag )
	+ \sum _{m=1}^{N^{\rm (mol)}}\sum _{j=1}^{N_m^{\rm (vib)}}\sum_k \hbar \Omega_{mjk}^{(\rm vib-diss)}(\hat r_{mjk}^\dag \hat b_{mj} 
	+ \hat r_{mjk} \hat b_{mj}^\dag ) \\
	+  \sum _{m=1}^{N^{\rm (mol)}}\sum _{j=1}^{N_m^{\rm (vib)}}\sum_k \hbar \Omega_{mjk}^{(\rm vib-deph)}\hat b_{mj}^\dag \hat b_{mj}(\hat c_{mjk}^\dag+\hat c_{mjk}) .
\end{multline}

One can write Heisenberg equations $d\hat A/dt=i\left[ \hat H^{\rm (total)}, \hat A\right]/\hbar$ for operators $\hat \sigma_m$ and $\hat b_{mj}$ and then eliminate environment degrees of freedom via Born-Markov approximation. It leads us to equations
\begin{multline}
	\frac{d}{dt}\hat \sigma_m=-(i\omega_m^{(\rm el)}+\gamma_{\bot })\hat \sigma_m-i\hat\sigma_m\sum_{j=1}^{N^{\rm (vib)}_m}\Lambda_{mj}\omega^{\rm (vib)}_{mj}(\hat b_{mj}+\hat b_{mj}^\dagger) \\
	+i\left(\Omega_m^{\rm (write)}(t)+\Omega_m^{\rm (read)}(t-\tau)\right)(2\hat \sigma_m^\dagger\hat\sigma_m-1)+\hat F_{\hat \sigma_m}(t),
\end{multline}
\begin{equation}
	\frac{d}{dt}\hat b_{mj}=-(i\omega_{mj}^{\rm (vib)}+\gamma_{mj}^{\rm (vib)})\hat b_{mj}-i\hat \sigma_m^\dagger \hat\sigma_m \Lambda_{mj}^2 + \hat F_{\hat b_{mj}}(t).
\end{equation}

Decay rates $\gamma_{\bot }$ and $\gamma_{mj}^{\rm (vib)}$ stand for relaxation of electronic and vibrational subsystems, respectively, and are determined by the environment. Noises $\hat F_{\hat \sigma_m}(t)$ and  $\hat F_{\hat b_{mj}}(t)$ are also determined by the environment and preserve commutation relations. For room temperatures $\hbar \omega_m^{\rm (el)}\gg k_BT$ so one can neglect  $\hat F_{\hat \sigma_m}(t)$. These equations can be approximately solved in the case of non-resonant excitation and weak exterlnal field which leads to Eq.~(\ref{sigma_St}) and~(\ref{sigma_aSt}).

\section{Electric field of the molecules} \label{molecules_field_derivation}
In this Appendix, we obtain an expression for the electric field of the Raman signal from the $m$th molecule at the input of the fiber.

\begin{figure} 
\includegraphics[width=1\linewidth]{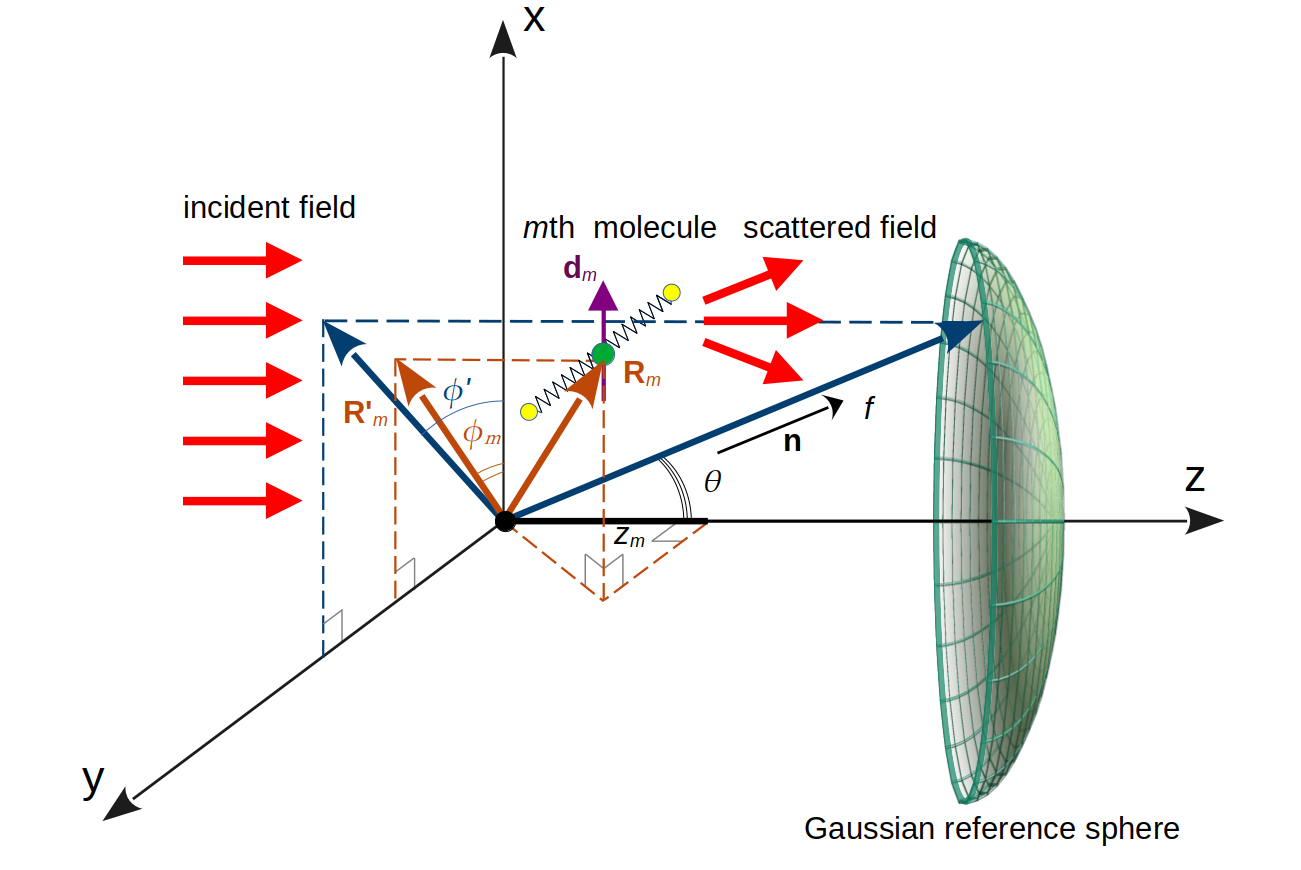}
\caption{
Geometry of the problem. 
The origin of the axes is in the focus of the first lense shown in Fig.~\ref{setup}.
} \label{molecules_in_focus}
\end{figure}

The electric field of the Stokes signal from the $m$th molecule at the input of the fiber is defined by this electic field at the Gaussian reference sphere of the radius $f$ centered at the focus of the first lens~\cite{hohenester2020nano} ({Fig.~\ref{molecules_in_focus}}). 
The Fourier component of this field at the frequency $\omega$ in the paraxial approximation is
\begin{equation}
\hat E_{m, \omega}^{\rm (St)} ({\bf r})=D(\omega)e^{i\omega t_0}\hat \sigma^{\rm (St)}_m(\omega)\int_0^{\theta_{\rm max}}d\theta \theta \int_0^{2 \pi}d\phi'e^{-i k r \sin\theta\cos{(\phi-\phi')}/M}e^{-ik \tilde n({\bf n}, {\bf R}_m)},
\end{equation}
where $\hat \sigma_m^{\rm (St)}(\omega)$ is the Fouerier transform of $\hat \sigma_m^{\rm (St)}(t)$ at the frequency $\omega$, $D(\omega)$ and $t_0$ depend on the parameters of the optical system, $\bf{r}$ is a 2D vector in the fiber input plane, $(r, \phi)$ are the corresponding polar coordinates, $k=\omega/c$, $M=f'/f$ (Fig.~\ref{setup}). 
Vector $\bf n$ is a unit vector and it is directed at angles ($\theta$, $\phi'$) from the focus of the first lens, ${\bf n}=(\sin{\theta}\cos{\phi'}, \sin{\theta}\sin{\phi'}, \cos{\theta})$. 
Let ${\bf R}_m={\bf R}_m'+{\bf{\hat z}}z_m$, where ${\bf R}_m'= (R_m'\cos\phi_m, R_m'\sin\phi_m, 0)$ and lies in the plane $xy$. 
It is easy to see that $e^{-ik \tilde n ({\bf n}, {\bf R}_m)}=e^{-ik \tilde n z_m\cos \theta}e^{-ik \tilde n R_m'\sin{\theta}\cos{(\phi_m-\phi')}}$, where $\phi_m$ is the angle between the vector ${\bf R}_m'$ and $x$ axis. 
Integrating over $\phi'$, we obtain 
\begin{equation}
\hat E_{m, \omega}^{\rm (St)} ({\bf r})=2\pi D(\omega)e^{i\omega t_0}\hat \sigma^{\rm (St)}_m(\omega)\int_0^{\theta_{\rm max}}d\theta \theta e^{-ik \tilde n z_m\cos \theta}J_0\left( \frac{k}{M}\left|{\bf r} + \tilde n M {\bf R}'_m \right|\sin\theta\right),
\end{equation}
where $J_0$ is the Bessel function of the zeroth kind.
The Stokes component of the scattered electric field from $m$th molecule in the time domain, $\hat E_m^{\rm (St)} ({\bf r}, t)=\int \hat E_{m, \omega}^{\rm (St)} ({\bf r})e^{-i\omega t}d\omega/2\pi$, is 
\begin{equation} \label{electric field St}
\hat E_m^{\rm (St)} ({\bf r}, t)=D(\omega^{\rm (St)})\int_0^{\theta_{\rm max}}d\theta \theta J_0\left( \frac{k^{\rm (St)}}{M}\left|{\bf r} + \tilde n M {\bf R}'_m \right|\sin\theta\right)\hat \sigma_m^{\rm (St)}\left(t-t_0+\frac{\tilde n z_m}{c}\cos \theta\right),
\end{equation}
where we used the quasi-monochromaticity of this component of the scattered light and introduced $k^{\rm (St)} = \omega^{\rm (St)}/c$.
Let $\hat \sigma_m^{\rm (St)}(t)=\hat{\tilde \sigma}_m^{\rm (St)}(t)e^{-i\omega^{\rm (St)}t}$, where $\hat{\tilde \sigma}_m^{\rm (St)}(t)$ changes slowly. 
In the paraxial approximation, $\cos \theta \approx 1-\theta^2/2$, $\sin \theta \approx \theta$, and
\begin{multline}
\hat \sigma_m^{\rm (St)}\left(t-t_0+\frac{\tilde n z_m}{c}\cos \theta\right)\approx \hat{\tilde \sigma}_m^{\rm (St)}\left(t-t_0+\frac{\tilde n z_m}{c}\right)e^{-i\omega^{\rm (St)}\left(t-t_0+\tilde n z_m/c\right)}e^{i\tilde n k^{\rm (St)}z_m\theta^2/2} \\
=\hat \sigma_m^{\rm (St)}\left(t-t_0+\frac{\tilde n z_m}{c}\right)e^{i\tilde n k^{\rm (St)}z_m\theta^2/2}.
\end{multline}
which allows us to transform Eq.~(\ref{electric field St}) 
\begin{equation} \label{Stokes field 1 molecule}
\hat E_m^{\rm (St)} ({\bf r}, t)=D^{\rm (St)}\hat \sigma_m^{\rm (St)}\left(t-t_0+\frac{\tilde n z_m}{c}\right)u^{\rm (St)}({\bf r}, {\bf R}_m), 
\end{equation}
where 
\begin{equation}
u^{\rm (St)}({\bf r}, {\bf R}_m)=\frac{(k^{\rm (St)})^2}{2 \pi M^2}\int_0^{\theta_{\rm max}}d\theta \theta J_0\left( \frac{k^{\rm (St)}}{M}\left|{\bf r} + \tilde n M {\bf R}'_m \right|\theta\right)e^{i\tilde n k^{\rm (St)}z_m\theta^2/2}.
\end{equation}

Here we considered the Stokes signal from $m$th molecule at the input of the fiber.
The derivation of the electric field for the anti-Stokes component of the scattered field is similar and straightforward. 
The result of the derivation is
\begin{equation} \label{anti-Stokes field 1 molecule}
\hat E_m^{\rm (aSt)} ({\bf r}, t)=D^{\rm (aSt)}\hat \sigma_m^{\rm (aSt)}\left(t-t_0+\frac{\tilde n z_m}{c}\right)u^{\rm (aSt)}({\bf r}, {\bf R}_m), 
\end{equation}
where 
\begin{equation}
u^{\rm (aSt)}({\bf r}, {\bf R}_m)=\frac{(k^{\rm (aSt)})^2}{2 \pi M^2}\int_0^{\theta_{\rm max}}d\theta \theta J_0\left( \frac{k^{\rm (aSt)}}{M}\left|{\bf r} + \tilde n M {\bf R}'_m \right|\theta\right)e^{i\tilde n k^{\rm (aSt)}z_m\theta^2/2},
\end{equation}
$k^{\rm (aSt)} = \omega^{\rm (aSt)}/c$, and $D^{\rm (St/aSt)}$ is defined in the main text after Eq.~(\ref{E}).

To obtain the input Stokes/anti-Stokes signal field, we sum Eq.~(\ref{Stokes field 1 molecule}) and Eq.~(\ref{anti-Stokes field 1 molecule}) over all molecules
\begin{equation}
\hat E_{\rm in}^{\rm (St/aSt)} ({\bf r}, t)=D^{\rm (St/aSt)}\sum_{m=1}^{N^{\rm (mol)}}\hat \sigma_m^{\rm (St/aSt)}\left(t-t_0+\frac{\tilde n z_m}{c}\right)u^{\rm (St/aSt)}({\bf r}, {\bf R}_m).
\end{equation}

\section{Derivation of (\ref{physical_meaning})} \label{pm_derivation}
We denote the numerator of the right-hand side of Eq.~(\ref{g2}) as $\tilde G^{(2)}_{\rm St,aSt}(\tau)$, we denote the denominator as $\tilde I^{\rm (St-aSt)}(\tau)$, so $\tilde g^{(2)}_{\rm St,aSt}(\tau)=\tilde G^{(2)}_{\rm St,aSt}(\tau)/\tilde I^{\rm (St-aSt)}(\tau)$. We transform $\tilde G^{(2)}_{\rm St,aSt}(\tau)$ using Eq.~(\ref{Output field}),
\begin{multline}
\tilde G^{(2)}_{\rm St,aSt}(\tau)
=
|A^{\rm (St)}|^2|A^{\rm (aSt)}|^2
\int d^2{\bf r}_1\int d^2{\bf r}_2\int\limits_{-\infty}^{+\infty} dt_1\int\limits_{-\infty}^{+\infty} dt_2\int\limits_{-\infty}^{+\infty}dt_1'\int\limits_{-\infty}^{+\infty}dt_2'\int\limits_{-\infty}^{+\infty}dt_3'\int\limits_{-\infty}^{+\infty}dt_4'
\\
\sum\limits_{k_1=1}^{N^{\rm (modes)}_{\rm LP}}\sum\limits_{k_2=1}^{N^{\rm (modes)}_{\rm LP}}
\sum\limits_{k_3=1}^{N^{\rm (modes)}_{\rm LP}}\sum\limits_{k_4=1}^{N^{\rm (modes)}_{\rm LP}}
f_{k_1}^{\rm (St)}({\bf r}_1)f_{k_2}^{\rm (aSt)}({\bf r}_2)
f_{k_3}^{\rm (aSt)}({\bf r}_2)f_{k_4}^{\rm (St)}({\bf r}_1)
G_{k_1}^{*}(t_1-t_1')G_{k_2}^{*}(t_2-t_2')\\
G_{k_3}(t_2-t_3')G_{k_4}(t_1-t_4') 
\left\langle 
\mathcal{T}_\rightarrow \{  \hat E_{k_1}^{{\rm (St)}\dag} (t_1') \hat E_{k_2}^{{\rm (aSt)}\dag} (t_2') \}
\mathcal{T}_\leftarrow  \{  \hat E_{k_3}^{\rm (aSt)} (t_3') \hat E_{k_4}^{\rm (St)} (t_4')  \}
\right\rangle
,
\end{multline}
where $G_k(\Delta t)=0$ at $\Delta t<0$. 
Taking into account the orthogonality of the modes, $\int d^2{\bf r}_1 f_{k_1}^{\rm (St)}({\bf r}_1)f_{k_4}^{\rm (St)}({\bf r}_1)=\delta_{k_1k_4}$ and $\int d^2{\bf r}_2 f_{k_2}^{\rm (aSt)}({\bf r}_2)f_{k_3}^{\rm (aSt)}({\bf r}_2)=\delta_{k_2k_3}$, discussed in the main text, and the identity $\int\limits_{-\infty}^{+\infty}dtG^{*}_{k}(t-t_1)G_{k}(t-t_2)=\delta(t_1-t_2)$, we obtain
\begin{multline} \label{N_approx}
\tilde G^{(2)}_{\rm St,aSt}(\tau)
=
|A^{\rm (St)}|^2|A^{\rm (aSt)}|^2
\\
\sum\limits_{k_1=1}^{N^{\rm (modes)}_{\rm LP}}\sum\limits_{k_2=1}^{N^{\rm (modes)}_{\rm LP}}\int\limits_{-\infty}^{+\infty} dt_1 \int\limits_{-\infty}^{+\infty} dt_2 
\left\langle 
\mathcal{T}_\rightarrow \{  \hat E_{k_1}^{{\rm (St)}\dag} (t_1) \hat E_{k_2}^{{\rm (aSt)}\dag} (t_2) \}
\mathcal{T}_\leftarrow  \{  \hat E_{k_2}^{\rm (aSt)} (t_2) \hat E_{k_1}^{\rm (St)} (t_1)  \} 
\right\rangle.
\end{multline}
From Eq.~(\ref{E}) and Eq.~(\ref{Stokes_mode}), it follows that $\hat E_k^{\rm (St)}$ and $\hat E_k^{\rm (aSt)}$ are linear combinations of the operators $\hat \sigma_{m}^{\rm (St/aSt)}$, which are uncorrelated for different $m$.
Taking into account this fact and that $N^{\rm (mol)}\gg 1$, we obtain 
\begin{multline} \label{G2 St aSt}
\tilde G^{(2)}_{\rm St,aSt}(\tau) 
\approx
|A^{\rm (St)}|^2|A^{\rm (aSt)}|^2
\\
\left[
\sum\limits_{k_1=1}^{N^{\rm (modes)}_{\rm LP}}\sum\limits_{k_2=1}^{N^{\rm (modes)}_{\rm LP}}
\int\limits_{-\infty}^{+\infty} dt_1 
\int\limits_{-\infty}^{+\infty} dt_2 
\langle \hat E_{k_1}^{\rm (St)\dag} (t_1) \hat E_{k_1}^{\rm (St)} (t_1) \rangle \langle \hat E_{k_2}^{\rm (aSt)\dag} (t_2) \hat E_{k_2}^{\rm (aSt)} (t_2) \rangle
\right.
\\
+
\sum\limits_{k_1=1}^{N^{\rm (modes)}_{\rm LP}}\sum\limits_{k_2=1}^{N^{\rm (modes)}_{\rm LP}}
\int\limits_{-\infty}^{+\infty} dt_2 
\int\limits_{-\infty}^{t_2} dt_1 
\langle \hat E_{k_1}^{{\rm (St)}\dag} (t_1) \hat E_{k_2}^{{\rm (aSt)}\dag} (t_2) \rangle \langle \hat E_{k_2}^{\rm (aSt)} (t_2) \hat E_{k_1}^{\rm (St)} (t_1) \rangle
\\
\left.
+
\sum\limits_{k_1=1}^{N^{\rm (modes)}_{\rm LP}}\sum\limits_{k_2=1}^{N^{\rm (modes)}_{\rm LP}}
\int\limits_{-\infty}^{+\infty} dt_1 
\int\limits_{-\infty}^{t_1} dt_2 
\langle \hat E_{k_2}^{{\rm (aSt)}\dag} (t_2) \hat E_{k_1}^{{\rm (St)}\dag} (t_1) \rangle \langle \hat E_{k_1}^{\rm (St)} (t_1) \hat E_{k_2}^{\rm (St)} (t_2) \rangle
\right].
\end{multline}
When deriving the equation above we also used that and anti-Stokes signals are hosted by thermal oscillations of molecules nuclei taking into account Eq.~(\ref{sigma_St}) and Eq.~(\ref{sigma_aSt}).

The transformation of the denominator of the right-hand side of Eq.~(\ref{g2}) is similar to the numerator and leads to expression
\begin{multline} \label{ISt-aSt}
\tilde I^{\rm (St-aSt)}(\tau)=
|A^{\rm (St)}|^2|A^{\rm (aSt)}|^2
\\
\sum\limits_{k_1=1}^{N^{\rm (modes)}_{\rm LP}}\sum\limits_{k_2=1}^{N^{\rm (modes)}_{\rm LP}}
\int\limits_{-\infty}^{+\infty} dt_1 
\int\limits_{-\infty}^{+\infty} dt_2 
\langle \hat E_{k_1}^{\rm (St)\dag} (t_1) \hat E_{k_1}^{\rm (St)} (t_1) \rangle \langle \hat E_{k_2}^{\rm (aSt)\dag} (t_2) \hat E_{k_2}^{\rm (aSt)} (t_2) \rangle.
\end{multline}

Substitution of Eq.~(\ref{G2 St aSt}) and~(\ref{ISt-aSt}) into Eq.~(\ref{g2}) leads to Eq.~(\ref{physical_meaning}).

\section{Pair correlations of the scattered electric field} \label{corr_derivation}
In this Appendix, we derive the pair correlations $ \langle \hat E^{\rm (St)\dag}_{k_1}(t_1) \hat E^{\rm (St)}_{k_2}(t_2) \rangle $, $ \langle \hat E^{\rm (aSt)\dag}_{k_1}(t_1) \hat E^{\rm (aSt)}_{k_2}(t_2) \rangle $, $ \langle \hat E^{\rm (St)}_{k_1}(t_1) \hat E^{\rm (aSt)}_{k_2}(t_2) \rangle $, and $ \langle \hat E^{\rm (aSt)}_{k_1}(t_1) \hat E^{\rm (St)}_{k_2}(t_2) \rangle $.

We consider the correlation $\langle \hat E_{k_1}^{{\rm (St)}} (t_1) \hat E_{k_2}^{{\rm (aSt)}} (t_2)\rangle$ in details.
Eq.~(\ref{Stokes_mode}) allows us to represent the correlation $\langle \hat E_{k_1}^{{\rm (St)}} (t_1) \hat E_{k_2}^{{\rm (aSt)}} (t_2)\rangle$ in the form
\begin{equation} \label{corr 2}
\langle \hat E_{k_1}^{{\rm (St)}} (t_1) \hat E_{k_2}^{{\rm (aSt)}} (t_2)\rangle=
\int d^2{\bf r}_1\int d^2{\bf r}_2
f_{k_1}^{\rm (St)}({\bf r}_1)f_{k_2}^{\rm (aSt)}({\bf r}_2)\langle 
\hat E^{\rm (St)}_{\rm in}({\bf r}_1,t_1) E^{\rm (aSt)}_{\rm in}({\bf r}_2,t_2)\rangle. 
\end{equation}
From Eq.~(\ref{E}), it follows 
\begin{multline}
\langle \hat E^{\rm (St)}_{\rm in}({\bf r}_1,t_1) \hat E^{\rm (aSt)}_{\rm in}({\bf r}_2,t_2)\rangle
=
D^{\rm (St)}D^{\rm (aSt)}\sum_{m=1}^{N^{\rm (mol)}}u^{(\rm St)}({\bf r}_1,{\bf R}_m)u^{(\rm aSt)}({\bf r}_2,{\bf R}_m) \\
\left\langle 
\hat \sigma_m^{(\rm St)}\left(t_1-t_0+\frac{\tilde n z_m}{c}\right)\hat\sigma_m^{(\rm aSt)}\left(t_2-t_0+\frac{\tilde n z_m}{c}\right) 
\right\rangle,
\end{multline}
where we use that $\hat \sigma_m^{\rm (St/aSt)}$ are uncorrelated for different $m$. 
Eq.~(\ref{sigma_St})--(\ref{sigma_aSt}) and the assumptions $\omega_{mj}^{\rm (vib)}\approx \omega^{\rm (vib)}$, $\omega_{m}^{\rm (el)}\approx \omega^{\rm (el)}$, $n_{mj}^{\rm (th)} \approx n^{\rm (th)}_{\rm vib}$ lead to 
\begin{multline} \label{full_double_corr}
\langle \hat E^{\rm (St)}_{\rm in}({\bf r}_1,t_1) \hat E^{\rm (aSt)}_{\rm in}({\bf r}_2,t_2)\rangle
=
n^{\rm (th)}_{\rm vib}D^{\rm (St)}_1 D^{\rm (aSt)}_1\sum_{m=1}^{N^{\rm (mol)}}\sum_{j=1}^{N_m^{\rm (vib)}}u^{(\rm St)}({\bf r}_1,{\bf R}_m)u^{(\rm aSt)}({\bf r}_2,{\bf R}_m) \\
\Lambda_{mj}^2e^{-\gamma_{mj}^{\rm (vib)}|t_1-t_2|}e^{i \omega_{mj}^{\rm (vib)}(t_1-t_2)}\Omega^{\rm (write)}_m\left(t_1-t_0+\frac{\tilde n z_m}{c}\right)
\Omega^{\rm (read)}_m\left(t_2-\tau-t_0+\frac{\tilde n z_m}{c}\right),
\end{multline}
where 
\begin{equation}
D^{\rm (St)}_1=\frac{D^{\rm (St)}\omega^{\rm (vib)}}{\left(\omega^{\rm (write)}-\omega^{\rm (vib)}-\omega^{\rm (el)} \right)\left(\omega^{\rm (write)}-\omega^{\rm (el)} \right)},
\end{equation}
\begin{equation}
D^{\rm (aSt)}_1=\frac{D^{\rm (aSt)}\omega^{\rm (vib)}}{\left(\omega^{\rm (read)}+\omega^{\rm (vib)}-\omega^{\rm (el)} \right)\left(\omega^{\rm (read)}-\omega^{\rm (el)} \right)}.
\end{equation}
Let $\mathcal{N}_p$ be a set of molecules of the $p$th type. 
Since the number of the molecules of each type is macroscopic, we can replace the sum over $m$ by the integral over the sample 
\begin{equation} \label{sum_and_integral}
\sum_{m=1}^{N^{\rm (mol)}}\sum_{j=1}^{N^{\rm (vib)}_m}
...
=
\sum_{p}\sum_{m\in \mathcal{N}_p}\sum_{j=1}^{N_p^{\rm (vib)}}
...
\approx 
\sum_{p} N_p \sum_{j=1}^{N_p^{\rm (vib)}} \int N({\bf R}_m) d^3{\bf R}_m
...
,
\end{equation}
where $N_p$ is the fraction of molecules of the $p$th type and $N({\bf R}_m)$ is the concentation of molecules.
We use Eq.~(\ref{Rabi frequency}) and Eq.~(\ref{sum_and_integral}) to transform Eq.~(\ref{full_double_corr})
\begin{multline} \label{corr 2 in}
\langle \hat E^{\rm (St)}_{\rm in}({\bf r}_1,t_1) \hat E^{\rm (aSt)}_{\rm in}({\bf r}_2,t_2)\rangle=
n^{\rm (th)}_{\rm vib}D^{\rm (St)}_2 D^{\rm (aSt)}_2
e^{-i(\omega^{\rm (write)}(t_1-t_0)+\omega^{\rm (read)}(t_2-t_0))} \\
F^{\rm (write)}(t_1-t_0)
F^{\rm (read)}(t_2-\tau-t_0)
\sum_{p} N_p \sum_{j=1}^{N_p^{\rm (vib)}} 
\Lambda_{pj}^2
e^{-\gamma_{pj}^{\rm (vib)}|t_1-t_2|}e^{i \omega_{pj}^{\rm (vib)}(t_1-t_2)}\\
\int d^3{\bf R}_m N({\bf R}_m) u^{(\rm St)}({\bf r}_1,{\bf R}_m)u^{(\rm aSt)}({\bf r}_2, {\bf R}_m)e^{-R'^2_m/w_0^2},
\end{multline}
where $D^{\rm (St)}_2=D^{\rm (St)}_1A^{\rm (write)}$ and $D^{\rm (aSt)}_2=D^{\rm (aSt)}_1A^{\rm (read)}e^{i\omega^{\rm (read)}\tau}$.
We substitute Eq.~(\ref{corr 2 in}) into Eq.~(\ref{corr 2}) and obtain
\begin{multline}
\langle \hat E_{k_1}^{{\rm (St)}} (t_1) \hat E_{k_2}^{{\rm (aSt)}} (t_2)\rangle=
n^{\rm (th)}_{\rm vib}D^{\rm (St)}_2 D^{\rm (aSt)}_2
e^{-i(\omega^{\rm (write)}(t_1-t_0)+\omega^{\rm (read)}(t_2-t_0))} \\
F^{\rm (write)}(t_1-t_0)
F^{\rm (read)}(t_2-\tau-t_0)
\sum_{p} N_p \sum_{j=1}^{N_p^{\rm (vib)}} 
\Lambda_{pj}^2
e^{-\gamma_{pj}^{\rm (vib)}|t_1-t_2|}e^{i \omega_{pj}^{\rm (vib)}(t_1-t_2)}\\
\int d^3{\bf R}_m  F^{(\rm St)}_{k_1}({\bf R}_m)F^{(\rm aSt)}_{k_2}({\bf R}_m)e^{-R_m'^2/w_0^2},
\end{multline}
where $F^{\rm (St/aSt)}_k({\bf R}_m)=\int d^2{\bf r}f^{\rm (St/aSt)}_k({\bf r})u^{\rm (St/aSt)}({\bf r}, {\bf R}_m)$. 
We define the scalar product $(F_1,F_2)=\int d^3{\bf R}_m N({\bf R}_m)  F^{*}_1({\bf R}_m)F_2({\bf R}_m)e^{-R_m'^2/w_0^2}$, and, finally, obtain
\begin{multline}
\langle \hat E_{k_1}^{{\rm (St)}} (t_1) \hat E_{k_2}^{{\rm (aSt)}} (t_2)\rangle=
n^{\rm (th)}_{\rm vib}D^{\rm (St)}_2 D^{\rm (aSt)}_2 
e^{-i(\omega^{\rm (write)}(t_1-t_0)+\omega^{\rm (read)}(t_2-t_0))} \\
F^{\rm (write)}(t_1-t_0)
F^{\rm (read)}(t_2-\tau-t_0)
(F^{\rm (St)*}_{k_1},F^{\rm (aSt)}_{k_2})
\sum_{m=1}^{N^{\rm (mol)}}\sum_{j=1}^{N_m^{\rm (vib)}} 
\Lambda_{mj}^2
e^{-\gamma_{mj}^{\rm (vib)}|t_1-t_2|}e^{i \omega_{mj}^{\rm (vib)}(t_1-t_2)}.
\end{multline}

The remaining two-point correlations are calculated similarly
\begin{multline}
\langle \hat E^{\rm (St)\dag}_{k_1}(t_1) \hat E^{\rm (St)}_{k_2}(t_2) \rangle
=(1+n^{\rm (th)}_{\rm vib})|D^{\rm (St)}_2|^2 e^{i\omega^{\rm (write)}(t_1-t_2)} \\
F^{\rm (write)}(t_1-t_0)
F^{\rm (write)}(t_2-t_0)
(F^{\rm (St)}_{k_1},F^{\rm (St)}_{k_2})
\sum_{m=1}^{N^{\rm (mol)}}\sum_{j=1}^{N_m^{\rm (vib)}} 
\Lambda_{mj}^2
e^{-\gamma_{mj}^{\rm (vib)}|t_1-t_2|}e^{-i \omega_{mj}^{\rm (vib)}(t_1-t_2)},
\end{multline}
\begin{multline}
\langle \hat E^{\rm (aSt)\dag}_{k_1}(t_1) \hat E^{\rm (aSt)}_{k_2}(t_2) \rangle
=n^{\rm (th)}_{\rm vib}|D^{\rm (aSt)}_2|^2 e^{i\omega^{\rm (read)}(t_1-t_2)} \\
F^{\rm (read)}(t_1-\tau-t_0)
F^{\rm (read)}(t_2-\tau-t_0)
(F^{\rm (aSt)}_{k_1},F^{\rm (aSt)}_{k_2})
\sum_{m=1}^{N^{\rm (mol)}}\sum_{j=1}^{N_m^{\rm (vib)}} 
\Lambda_{mj}^2
e^{-\gamma_{mj}^{\rm (vib)}|t_1-t_2|}e^{i \omega_{mj}^{\rm (vib)}(t_1-t_2)},
\end{multline}

\begin{multline}
\langle \hat E^{\rm (aSt)}_{k_1}(t_1) \hat E^{\rm (St)}_{k_2}(t_2) \rangle
=
(1+n^{\rm (th)}_{\rm vib})D^{\rm (aSt)}_2 D^{\rm (St)}_2
e^{-i(\omega^{\rm (read)}(t_1-t_0)+\omega^{\rm (write)}(t_2-t_0))} \\
F^{\rm (read)}(t_1-\tau-t_0)
F^{\rm (write)}(t_2-t_0)
(F^{\rm (aSt)*}_{k_1},F^{\rm (St)}_{k_2})
\sum_{m=1}^{N^{\rm (mol)}}\sum_{j=1}^{N_m^{\rm (vib)}} 
\Lambda_{mj}^2
e^{-\gamma_{mj}^{\rm (vib)}|t_1-t_2|}e^{-i \omega_{mj}^{\rm (vib)}(t_1-t_2)}.
\end{multline}

\section{Expressions for $I_{m_1j_1m_2j_2}^{\rm (\alpha,\beta)}$} \label{I_coeff}
\begin{multline}
	I_{m_1j_1m_2j_2}^{\rm ({\rm St},{\rm aSt})}=\int_{-\infty}^{+\infty}dt_2\int_{-\infty}^{t_2}dt_1[F^{\rm (write)}(t_1)F^{\rm (read)}(t_2-\tau)]^2
	\\
	e^{-i(\omega_{m_1j_1}^{\rm  (vib)}-\omega_{m_2j_2}^{\rm  (vib)})(t_1-t_2)}e^{-(\gamma_{m_1j_1}^{\rm  (vib)}+\gamma_{m_2j_2}^{\rm  (vib)})|t_1-t_2|},
\end{multline}
\begin{multline}
	I_{m_1j_1m_2j_2}^{\rm ({\rm aSt},{\rm St})}=\int_{-\infty}^{+\infty}dt_1\int_{-\infty}^{t_1}dt_2[F^{\rm (write)}(t_1)F^{\rm (read)}(t_2-\tau)]^2
	\\
	e^{-i(\omega_{m_1j_1}^{\rm  (vib)}-\omega_{m_2j_2}^{\rm  (vib)})(t_1-t_2)}e^{-(\gamma_{m_1j_1}^{\rm  (vib)}+\gamma_{m_2j_2}^{\rm  (vib)})|t_1-t_2|},
\end{multline}
\begin{multline}
	I_{m_1j_1m_2j_2}^{\rm (St,St)}
	=
	\int\limits_{-\infty}^{+\infty} dt_1
	\int\limits_{-\infty}^{+\infty} dt_2
	[F^{\rm (write)}(t_1)F^{\rm (write)}(t_2)]^2 \\
	e^{-i(\omega_{m_1j_1}^{\rm  (vib)}-\omega_{m_2j_2}^{\rm  (vib)})(t_1-t_2)}e^{-(\gamma_{m_1j_1}^{\rm  (vib)}+\gamma_{m_2j_2}^{\rm  (vib)})|t_1-t_2|},
\end{multline}
\begin{multline}
	I_{m_1j_1m_2j_2}^{\rm (aSt,aSt)}
	=
	\int\limits_{-\infty}^{+\infty} dt_1
	\int\limits_{-\infty}^{+\infty} dt_2
	[F^{\rm (read)}(t_1)F^{\rm (read)}(t_2)]^2 \\
	e^{-i(\omega_{m_1j_1}^{\rm  (vib)}-\omega_{m_2j_2}^{\rm  (vib)})(t_1-t_2)}e^{-(\gamma_{m_1j_1}^{\rm  (vib)}+\gamma_{m_2j_2}^{\rm  (vib)})|t_1-t_2|}.
\end{multline}

\bibliography{fiber_and_correlations}

\end{document}